%
% the following is to use blackboard bold fonts --
\let\useblackboard=\iftrue
%
% activate this if you don't have them.
%\let\useblackboard=\iffalse
%
% You might also need to remove this line.
\newfam\black
\input harvmac.tex
\def\Title#1#2{\rightline{#1}
\ifx\answ\bigans\nopagenumbers\pageno0\vskip1in%
\baselineskip 15pt plus 1pt minus 1pt
\else%\special{papersize=11in,8.5in}%
\def\listrefs{\footatend\vskip 1in\immediate\closeout\rfile\writestoppt
\baselineskip=14pt\centerline{{\bf References}}\bigskip{\frenchspacing%
\parindent=20pt\escapechar=` \input
refs.tmp\vfill\eject}\nonfrenchspacing}
\pageno1\vskip.8in\fi \centerline{\titlefont #2}\vskip .5in}

scaled\magstep3
 
scaled\magstep3
\ifx\answ\bigans\def\tcbreak#1{}\else\def\tcbreak#1{\cr&{#1}}\fi
\useblackboard
\message{If you do not have msbm (blackboard bold) fonts,}
\message{change the option at the top of the tex file.}
\font\blackboard=msbm10 scaled \magstep1
\font\blackboards=msbm7
\font\blackboardss=msbm5
%\newfam\black
\textfont\black=\blackboard
\scriptfont\black=\blackboards
\scriptscriptfont\black=\blackboardss

\else

\fi
% *************************************
%
\def\yboxit#1#2{\vbox{\hrule height #1 \hbox{\vrule width #1
\vbox{#2}\vrule width #1 }\hrule height #1 }}
\def\fillbox#1{\hbox to #1{\vbox to #1{\vfil}\hfil}}
\def\ybox{{\lower 1.3pt \yboxit{0.4pt}{\fillbox{8pt}}\hskip-0.2pt}}

\def\comments#1{}

\def\CM{{\cal M}}
\def\CN{{\cal N}}

\def\CV{{\cal V}}

\def\II{\relax{I\kern-.07em I}}

\def\IZ{\relax\ifmmode\mathchoice
{\hbox{\cmss Z\kern-.4em Z}}{\hbox{\cmss Z\kern-.4em Z}}
{\lower.9pt\hbox{\cmsss Z\kern-.4em Z}}
{\lower1.2pt\hbox{\cmsss Z\kern-.4em Z}}\else{\cmss Z\kern-.4em
Z}\fi}
\def\IB{\relax{\rm I\kern-.18em B}}
\def\IC{{\relax\hbox{$\inbar\kern-.3em{\rm C}$}}}
\def\ID{\relax{\rm I\kern-.18em D}}
\def\IE{\relax{\rm I\kern-.18em E}}
\def\IF{\relax{\rm I\kern-.18em F}}
\def\IG{\relax\hbox{$\inbar\kern-.3em{\rm G}$}}
\def\IGa{\relax\hbox{${\rm I}\kern-.18em\Gamma$}}
\def\IH{\relax{\rm I\kern-.18em H}}
\def\II{\relax{\rm I\kern-.18em I}}
\def\IK{\relax{\rm I\kern-.18em K}}
\def\IP{\relax{\rm I\kern-.18em P}}
%\def\IX{\relax{\rm X\kern-.01em X}}
%this doesn't work

\font\cmss=cmss10 \font\cmsss=cmss10 at 7pt
\def\IR{\relax{\rm I\kern-.18em R}}

\def\bX{\bf X}
\def\bS{\bf S}
\def\bZ{\bf Z}
\def\bC{\bf C}

\Title{ \vbox{\baselineskip12pt\hbox{RU-96-68, hep-th/9608086}}}
{\vbox{
\centerline{Hypermultiplet Moduli Space}
\centerline{and String Compactification to Three Dimensions}}}
\centerline{Nathan Seiberg and Stephen Shenker}
\smallskip
\smallskip
\centerline{Department of Physics and Astronomy}
\centerline{Rutgers University }
\centerline{Piscataway, NJ 08855-0849}
\centerline{\tt seiberg@physics.rutgers.edu,
shenker@physics.rutgers.edu} 
\bigskip
\bigskip
\noindent
We study the hypermultiplet moduli space of the type II string
compactified on a Calabi-Yau space $\bX$.  We do this by using IIA/IIB
duality in a compactification of the same theory on $\bX\times \bS ^1$
and by using recent results on three dimensional field theory.
%\draftmode
\Date{August 1996}

The moduli of the type II theory compactified on a Calabi-Yau space $X$
are in hypermultiplets and vector multiplets of $\CN=2$ supersymmetry.
The moduli space has a product structure\foot{To get such
a product structure one might need to consider a multiple
cover of $\CV$ or $\CH$.}
$\CH \times \CV$.  $\CH $ is a
quaternionic manifold and $\CV$ is a special Kahler manifold.  Since the
dilaton is in a hypermultiplet, it appears only in $\CH$ and not in
$\CV$.  Therefore, $\CV $ is determined exactly in the classical theory.
$\CH$ on the other hand can receive quantum corrections.

Strominger's resolution of the conifold singularity
\ref\strominger{A. Strominger, ``Massless Black Holes and Conifolds in
String Theory,'' Nucl.Phys {\bf B451} (1995) 97, hep-th/9504090.}
left open the question of the behavior of
$\CH$ near a conifold point and in particular the question of
whether the space is singular or not.  It was suggested in
\ref\bbs{K. Becker, M. Becker and A. Strominger, ``Fivebranes,
Membranes and Non-Perturbative String Theory,'' Nucl.Phys. {\bf B456}
(1995) 130, hep-th/9507158.}
\nref\shenker{S. Shenker, ``Another length Scale in String Theory?''
hep-th/9509132.}%
that non-perturbative effects -- string instantons -- smooth out the
singularity.  It became immediately clear \bbs\ that this
question is intimately related to the behavior of the theory after
compactification on another circle to three dimensions \refs{\bbs,
\shenker}. 

Recently the compactification of such $\CN=2$ field theories {}from four
to three dimensions was analyzed
\ref\seiwit{N. Seiberg and N. Witten, ``Gauge Dynamics and
Compactification to Three Dimensions,'' hep-th/9607163.}.
The Coulomb branch of the moduli space of the four dimensional theory,
$\CV$, becomes in the three dimensional theory a hyper-Kahler manifold
$\tilde \CV$.  Consider for simplicity the case where $\CV$ is one
complex dimensional.  Then it is a base of an $SL(2,\bZ)$ vector bundle
\ref\sw{N. Seiberg and E. Witten, ``Electric-Magnetic Duality, Monopole
Condensation, And Confinement In $N=2$ Supersymmetric Yang-Mills
Theory,'' Nucl. Phys. {\bf B426} (1994) 19, hep-th/9407087; ``Monopoles,
Duality, And Chiral Symmetry Breaking In $N=2$ Supersymmetric QCD,''
Nucl. Phys. {\bf B431}  (1994)  484, hep-th/9408099.}.
$\tilde \CV$ is the four real dimensional vector bundle with the same
complex structure.  The Kahler form of the fiber scales as the inverse of
the radius $R$ of the compactification and that of the base scales as 
$R$ so 
that the volume form is independent of the radius.  Equivalently, we can
rescale the entire space and have a fiber whose volume is $1 \over R^2$.

Applying this to string theory, consider compactifying the type IIA
theory on $\bX$ with a moduli space 
\eqn\modfa{\CM^{(4)}_A= \CH_A \times \CV_A .}
Upon further compactification to three
dimensions on $\bS^1$ of radius 
$R_A$ the moduli space becomes\foot{We are not sure that the moduli
space is globally such a product (even if we consider multiple covers of
$\CH_A$ and $ \tilde \CV_A$).  Below we will need this fact only
locally where field theory is valid.  Then, the product structure
follows {}from the fact that two different $SU(2)$'s rotate the complex
structures of the two factors.}
\eqn\moda{\CM^{(3)}_A= \CH_A \times \tilde \CV_A.}
Locally, and in particular near conifold singularities, the field
theoretic analysis of \seiwit\ applies.  The volume of the fiber torus
is $1 \over M_p^2R_A^2$ where $M_p$ is the four dimensional Planck mass.
In terms of the IIA string coupling $\lambda_A$ and the string scale
$M_s$ it is
\eqn\toruvol{\left({\lambda_A \over M_sR_A}\right)^2.}

Now perform a T-duality transformation on the circle.  This leads to the
IIB theory compactified on $\bX \times \bS^1$ with radius and coupling 
\eqn\brc{\eqalign{R_B=&{1 \over M_s^2 R_A} \cr
\lambda_B=&{\lambda_A \over M_sR_A} . \cr}}
The moduli space of the IIB theory is
\eqn\modb{\CM^{(3)}_B= \CH_B \times \tilde \CV_B.}
By T-duality $\CM^{(3)}_A= \CM^{(3)}_B$ but
\eqn\modrel{\eqalign{\CH_B =& \tilde \CV_A \cr
\CH_A =& \tilde \CV_B .\cr}}

For the three dimensional theory a convenient coordinate system 
on the space of $\lambda_A$ and $R_A$ is
$\lambda_A$ and $\lambda_B={\lambda_A \over M_sR_A}$.  $\CH_A = \tilde
\CV_B$ depends only on $\lambda_A$ and $\CH_B = \tilde \CV_A$ depends
only on $\lambda_B$. 

As $R_A$ goes to zero, $R_B$ goes to infinity and the IIB theory becomes
four dimensional with moduli space 
\eqn\modfob{\CM^{(4)}_B= \CH_B \times \CV_B}
where $\tilde \CV_B$ is related to $\CV_B$ as explained above.
Therefore, all these moduli spaces are uniquely determined by $\CV_A$
and $\CV_B$.  In particular, $\CH_B$ is easily constructed out of
$\CV_A$ where the volume of the fiber is $\lambda_B^2$.

Using this information we can easily conclude a few facts about the
singularities in $\CH_B$.  Near a simple conifold, $\CV_A$ has a
singularity associated with a single massive hypermultiplet becoming
massless \strominger.  The corresponding three dimensional theory is
$U(1)$ with one electron whose moduli space $\tilde \CV_A$ is smooth
\nref\seibergt{N. Seiberg, ``IR Dynamics on Branes and Space-Time
Geometry,'' hep-th/9606017.}%
\refs{\seibergt,\seiwit}.  Therefore, $\CH_B$ at that point is smooth.
If $N$ massive hypermultiplets become massless at a point in $\CV_A$,
$\tilde \CV_A$ has an $A_{N-1}$ singularity \refs{\seibergt,\seiwit} -- it
is locally ${\bC}^2/{\bZ}_N$.  Therefore, $\CH_B$ should also have such a
singularity at that point.  More complicated conifolds like those
discussed in
\ref\KMP{A. Klemm and P. Mayr, ``Strong Coupling Symmetries and
Nonabelian Gauge Symmetries in N=2 String Theory,'' hep-th/9601014;
S. Katz, D. Morrison, and R. Plesser, ``Enhanced Gauge Symmetry in Type
II String Theory,'' hep-th/9601108.} 
can be analyzed in a similar way.

Recently 
\ref\ov{B.R. Greene, D.R. Morrison and C. Vafa, ``A Geometric
Realization of Confinement,'' hep-th/9608039; H. Ooguri and C. Vafa,
``Summing up D-Instantons,'' hep-th/9608079.} 
some of these results have been derived {}from a different point of
view. The instanton corrections they compute have a simple
interpretation in the four to three dimensional language.  They are
simply the world line windings of the light hypermultiplet around the
$\bS^1$ \shenker. 

To see this, we note that the metric for the $U(1)$ field theories with
$N$ hypermultiplets is one loop exact.  This fact can be derived as
follows.  If the gauge theory is Abelian, there are no magnetic
monopoles in four dimensions and there are no instantons in the three
dimensional theory.  Therefore, there is no need to perform a duality
transformation in the three dimensional Lagrangian.  Writing it in terms
of the $U(1)$ gauge field, an argument similar to
the one in four dimensions shows that only one loop corrections are
possible.  Therefore, $1/e^2(z) = 1/e_0^2+ N I(z)$ where $I(z)$ is the
one loop charge renormalization of one hypermultiplet of mass
${|z| \over \lambda}$
in the space ${\bf R}^3 \times \bS ^1$
\eqn\iz{
I(z)  \sim {1 \over R }\sum_{n} \int d^3 p 
{1 \over{(p^2 + ({{n} \over
{R}})^2 + ({|z| \over \lambda})^2)^2}} +const. \sim 
\sum_{n} {1 \over \sqrt{n^2+({|z|R\over \lambda})^2}} + const. }
For simplicity we set other scalar vevs to zero.  The (infinite)
additive constant in \iz\ corresponds to renormalization of $1/e_0^2$.
The function appearing in the metric for the dualized photon is
$e^2(z)={e_0^2\over{1+e_0^2 N I(z)}}$ and appears to have all order
corrections.  This is the origin of the positive mass Taub-NUT factors
found in \seiwit .  So all instanton effects come {}from the $R$
dependence of $I(z)$. For large $R$ this dependence comes {}from the
world lines of particles of mass ${|z| \over
\lambda}$ winding around the circle $l$ times,
producing an effect of weight
$\exp ({{-2 \pi R l |z|}\over \lambda})$.  These effects can be isolated
by Poisson resummation of \iz.  If \iz\ is cut off at mass $M$,  
e.g., by  a heavy
Pauli-Villars hypermultiplet, 
there will be additional $R$ dependence
of weight $\exp ({-2 \pi R l M})$ that signals the cutoff scale
\shenker .

\medskip
\centerline{\bf Acknowledgements}
This work was supported in part by DOE grant DE-FG02-96ER40559.

\listrefs
\end